\documentclass[prl,preprint]{revtex4}
\usepackage{graphicx}
\usepackage{dcolumn}
\usepackage{bm}
\begin{document}
\title {Tracking random walk of individual domain walls in cylindrical
nanomagnets with resistance noise}
\author{Amrita Singh\footnote[1]{electronic mail:amrita@physics.iisc.ernet.in}}
\author{Soumik Mukhopadhyay\footnote[2]{electronic mail:soumik@physics.iisc.ernet.in}}
\author{Arindam Ghosh}
\affiliation{Department of Physics, Indian Institute of Science, Bangalore 560 012, India}

\begin{abstract}

The stochasticity of domain wall (DW) motion in magnetic nanowires has been
probed by measuring slow fluctuations, or noise, in electrical resistance at
small magnetic fields. By controlled injection of DWs into isolated cylindrical
nanowires of nickel, we have been able to track the motion of the DWs between
the electrical leads by discrete steps in the resistance. Closer inspection of
the time-dependence of noise reveals a diffusive random walk of the DWs with an
universal kinetic exponent. Our experiments outline a method with which
electrical resistance is able to detect the kinetic state of the DWs inside the
nanowires, which can be useful in DW-based memory designs.

\end{abstract}


\maketitle

Domain wall (DW) dynamics in ferromagnetic nanowires has received intense
interest in recent years due to their potential application in a novel class of
memory devices, as well as from the viewpoint of exciting fundamental
physics~\cite{Parkin_Science_2008,Beach_JMM_2008,Allwood_Science}. The process
of DW propagation, driven either by magnetic field ($H$) or by electric
current, is intimately connected to the magnetization reversal mechanism, and
hence influenced by the geometry and magnetic anisotropy properties, as well as
intrinsic disorder within the nanowires that act as pinning centers. A
consequence of this is the non-deterministic, or stochastic, kinetics of the
DWs which is a subject of great fundamental and technological
importance~\cite{Parkin_Science_2008,Im_PRL_2008,Im_PRL_2009,Yamaguchi_PRL_2004,
Meier_PRL_2007,Durin_PRL_2000,Jourdan_PRB_2007}. Randomness associated with the
magnetization reversal process is the major source of stochasticity, and a
common mechanism involves random walk of DWs through Barkhausen avalanches,
where the DWs are treated as particles undergoing Brownian motion according to
the Langevin equation~\cite{ABBM_JAP_1990}. Such a mechanism has been
established in magnetic thin films, where the Barkhausen statistics reflects in
universal scaling exponents in the avalanche size/duration distribution
functions~\cite{Im_PRL_2008,Meier_PRL_2007}. The viscous DW flow has also been
observed in Mn-doped semiconductor epilayers~\cite{Tang_PRB_2006}, but a clear
signature of $H$-driven random walk of DWs in magnetic nanowires has not been
observed experimentally. This issue is of particular interest in the case
cylindrical magnetic nanowires, where the DWs behave as ``massless'' particles
with zero kinetic energy~\cite{Yan_PRL_2010}, and hence whether the general
mechanisms of stochastic kinetics can at all be relevant in this case is
unknown. In this work, we provide the first unambiguous evidence of diffusive
random walk of DWs in cylindrical high-aspect ratio magnetic nanowires by
tracking their motion at small $H$ (just above the depinning field). We show
that not only the stochasticity in DW kinteics within magnetic nanowires can be
described by Brownian diffusion with universal exponents, but the nature of the
stochasticity can be employed to probe the kinetic state of the DWs themselves.

Quantifying stochasticity with conventional probes such as the Kerr effect,
X-ray/electron or force microscopy, involve analyzing magnetization burst size,
variations in DW displacement or the depinning fields etc, where the
sensitivity to the evolution in the DW motion in time domain can be limited.
Time of flight probing, for example in the context of first time
arrival~\cite{Im_PRL_2008}, or planar Hall
effect~\cite{Tang_PRB_2006,Attane_PRL_2006,Nowak_Nowak_2009}, have been useful
in locating DWs or measure their average velocity between spatially separated
probes. Here we have adopted a different route, and measured the low-frequency
fluctuations in longitudinal electrical resistance ($R$) of magnetic nanowires
at small $H$ above the depinning threshold. In disordered metallic systems
these fluctuations, often known as $1/f$-noise, are extremely sensitive to slow
relaxation of defects (dislocations, cluster of point defects etc). Random
movement of the scatterers, even at a scale $\sim$ Fermi wavelength ($\lambda_F
\lesssim 1$~nm), change the interference pattern of coherently back-scattered
electrons, reflecting in the time-dependent fluctuations in the
resistivity~\cite{Pelz_PRB_1987}. In magnetic systems, the DWs themselves can
act as scatterers of spin-polarized conduction electrons, and modify the $R$ of
the nanowires. This can occur either through direct reflection when DW width
$\Delta \sim \lambda_F$~\cite{Berger_PRB_2006}, or by spin-dependent scattering
of electrons by the disorder inside the
DWs~\cite{Levy_PRL_1997,Gorkom_PRL_1999}. Recently, the fluctuations in $R$ in
different forms of nano-magnetic structures have been associated with the
motion of DWs~\cite{Cucchiara_APL_2009,Singh_APL_2009}, although the details of
the time dependence of $R$ due motion of individual DWs remain unexplored.

We have used nickel nanowires that were electrochemically grown inside anodic
alumina templates - a well-characterized system in the context of magnetic
storage~\cite{Wernsdorfer_PRL_1996,Nielsch_APL_2001,Paulus_JMM_2001}. We have
used nanowires of average diameter $\approx 200$~nm, where strong shape
anisotropy (aspect ratio $> 100$) aligns the easy-axis of magnetization along
the long axis of the nanowires. Details of the growth process and structural
characterization can be found elsewhere~\cite{Singh_APL_2009}. Following the growth, free standing nanowires
were obtained by dissolving the alumina template in $2$ M $NaOH$ solution. Nanowires were then drop-casted on
flat silicon oxide substrates, after which
electron-beam lithography was used to form
Ti/Au contact pads on the nanowire for electrical measurements. SEM micrograph of
the device used for the present experiments
appears in Fig.~1a where the length ($L$) of the nanowire between the voltage
probes (indicated by $V+$ and $V-$) was $\approx 4.5 \mu$~m. Edge roughness,
and also the branching/clusters attached at the end of the nanowire, reduce the
DW nucleation barrier substantially~\cite{Wernsdorfer_PRL_1996}. To measure
small changes in $R$, a dynamically balanced ac Wheatstone bridge arrangement was used (excitation frequency of $226$ Hz). [see Ref.~\cite{Singh_APL_2009} and
~\cite{Chandni_Acta_2010} for more details on noise measurements.] All
measurements were performed with a very low excitation current density
($\lesssim 10^{7}$~A/m$^2$) to avoid heating, electromigration, or the
spin-torque effect. The background fluctuations consisted mainly of Nyquist
noise, and the resolution to change in $R$ was $\sim 10$~ppm. Fig.~1b shows the
magnetoresistance curves at three different angles ($\theta$) between $H$ and
the electric current density (nanowire long axis). The nanowire exhibits
anisotropic magnetoresistance (AMR) where the switching field $H_{sw}$,
identified by the dip in the AMR, increases continuously from $\approx 250$~Oe at
at $\theta = 0$ to the maximum of $\approx 850$~Oe at $\theta = 90^\circ$. This
indicates the magnetization reversal to occur via curling mode as expected in
nickel nanowires with diameter $\gtrsim 45$~nm~\cite{Nielsch_APL_2001}.

To find signature of the DWs, expected to be of vortex type in our
case~\cite{Beach_JMM_2008}, we have recorded $R$ as a function of time at fixed
values of $H$ applied parallel to the nanowire axis. $H$ was increased
monotonically in small steps, starting from $H = 0$, and at every step time
dependence of $R$ was measured over $\approx 60$~min. In Fig~2, time series
recorded at four different $H$ are shown. At very low $H$ ($\ll 1$~Oe) the
fluctuations in $R$ are featureless with a power spectral density (PSD) of
noise, $S_R/R^2 \propto f^{-\alpha}$, where $\alpha \approx 1$ (also see
Fig.~3a). At $H \gtrsim 1$~Oe, discrete multi-level states in $R$ appear as a
function of time as shown in the two lower panels of Fig.~2 for $H = 1.5$~Oe
and 3~Oe. Further increase in $H$ somewhat obscures the visibility of the
multi-level states, which disappear completely for $H \gtrsim H_{sw}$ (time
series not shown). The same sequence was repeated over many magnetization
cycles, indicating the phenomena to be due to application of $H$, and not due
to relaxation of internal disorder driven by temperature or electric current.

Before analyzing the time-dependence of the fluctuations, we address the origin
of the discrete jumps in $R$ at $H \gtrsim 1$~Oe. In all cases the jumps
involve {\it increase} in $R$ from its base value $R_0$ ($\approx 4.0~\Omega$
at room temperature) by $\Delta R$ or $2\Delta R$, where $\Delta R \approx
3$~m$\Omega$. Since $H$ is kept fixed, AMR or Lorentz contributions to $R$ do
not change, and hence a natural explanation involves the DWs, which nucleate at
the defect sites and travel intermittently across the voltage probes.
Increasing $R$ by $\Delta R$ and $2\Delta R$ then corresponds to fitting a
domain partially (one DW) or fully (two DWs) between the voltage probes,
respectively. Indeed, the positive correction $\Delta R$ can be quantitatively
understood from the Levy-Zhang model of spin-mixing due to disorder scattering
inside the DWs~\cite{Levy_PRL_1997}, which estimates the fractional change in
$R$ from the inclusion of a single DW between the voltage probes as
$(\Delta/L)\times[1+\xi^2(\rho^\uparrow -
\rho^\downarrow)^2/5\rho^\uparrow\rho^\downarrow] \approx 0.7\%$. Here
$\rho^\uparrow$ and $\rho^\downarrow$ correspond to resistivities of the up and
down spin channels respectively with $\rho^\uparrow/\rho^\downarrow \approx 3$
in nickel, $\xi = \pi\hbar^2k_F/4m_eJ\Delta \approx 1$ with Fermi  wave vector
$k_F = 1.5\times10^{10}$~m$^{-1}$, nickel exchange energy $J =
4.46\times10^{-21}$~J, and DW width $\Delta = 24$~nm~\cite{Paulus_JMM_2001}.
Experimentally, we find $\Delta R/R_{4.2K} \approx 0.4 - 0.5\%$ which agrees
with the expected DW contribution within a factor of two (we used the low
temperature residual resistance $R_{4.2K} \approx 0.8~\Omega$ of the nanowire
to eliminate the phonon contribution). Occasionally, the time series at larger
$H$ shows direct jumps of $2\Delta R$ which could be due to nucleation of
domains within the region between the voltage probes. In the time domain, the
jumps did not show any regular pattern or sequence, which prompted us to focus
on the frequency domain through power spectral analysis.

The PSD of the fluctuations in $R$ over the entire ($\sim$ hour long) time
series was found to vary as $S_R/R^2 = A_R/f^\alpha$ (Fig.~3a), where both
noise amplitude ($A_R$) and $\alpha$ depend strongly (and non-monotonically) on
$H$ (Figs.~3b and 3c). Three regimes can be clearly identified, and understood
in term of the DWs: (1) At $H < 1$~Oe, which we can now identify as the
depinning field, the noise magnitude (expressed as relative variance $\delta
R^2/R^2 = \int (S_R/R^2)df$ in Fig.~3c) is low and $\alpha$ is $\approx 1 -
1.2$ (Fig.~3b). This $H$-independent background noise arises from slow
relaxation of disorder (such as dislocations, vacancy clusters etc.), which has
a PSD $\sim 1/f^\alpha$, where $\alpha \approx 1$, due to the broad
distribution of associated time scales. (2) For intermediate $H$ (1~Oe$< H
\lesssim H_{sw}$), we identify a sharp increase in both $\delta R^2/R^2$ and
$\alpha$ ($\sim 1.4 - 1.7$). We can understand this with two parallel
mechanisms. First, random fluctuations in the number of DWs between the voltage
probes lead to PSD $\sim 1/f^\alpha$ with $\alpha \sim 2$, and second, the
fluctuations in $R$ generated by any given DW during its flight between the
probes. The latter causes $R$ to fluctuate in a given state, and embodies the
stochasticity of DW propagation which will be treated separately. (3) Finally,
for $H > H_{sw}$ the number of domains diminish, and both $\delta R^2/R^2$ and
$\alpha$ return to their zero-field background values.

Can the stochasticity in DW propagation be extracted from the kinetics of
resistance noise? To answer this we return to Fig.~2, and focus on the
fluctuations only in the $R = R_0+2\Delta R$ state, which would correspond to
one propagating domain ({\it i.e.} two DWs) between the voltage probes. For a
preliminary time-of-flight analysis, we note that the time ($\tau_H$) that the
system stays in this state corresponds to the time the domain takes to travel
from one voltage probe to the other. In Fig.~4 two histograms of $\tau_H$
obtained at $H = 1.5$~Oe and 3~Oe are shown. Clearly, the modal magnitude of
$\tau_H$ decreases with increasing $H$, due to increase in the DW velocity
(inset). Two important points are to be noted here: (1) The width of the
velocity distribution deceases with increasing $H$, which can be attributed to
the $H$-induced reduction in the effective propagation barrier that suppresses
(lower) part of the barrier energy distribution. (2) Secondly, the typical
velocity is about five orders of magnitude lower than thin film-based magnetic
nanostrips, which can be understood from the suppression of DW mobility ($\sim
\Delta^{4.4}$) at greatly reduced DW width in nanowires of cylindrical cross
section~\cite{Berger_PRB_2006}.

The PSD of noise in the high resistance state (spanning over $\tau_H$) shows a
strikingly universal behavior. At all $\tau_H$ segments (see typical time
traces in Fig.~5a), the PSDs vary as $S_R/R^2 \sim 1/f^\alpha$, where $\alpha =
1.5\pm0.05$ at both $H = 1.5$~Oe and 3~Oe (Fig.~5b and 5c, respectively). This
is distinctly different in the low resistance states, where $\alpha$ was found
to be $\approx 1.0 - 1.2$ (not shown). In disordered metals, which is not
undergoing plastic deformation or any structural phase
transitions~\cite{Chandni_PRL_2009}, the observation of $\alpha = 1.5$ in
resistance noise signifies diffusion or random walk of the charge scatterers,
such as dislocations, vacancy/interstitial clusters
etc~\cite{Fourcade_PRB_1986,Ghosh_JPHYSD_1997}. The exponent is universal in
the sense that it is a direct outcome of the fluctuation-dissipation theorem
for a system in thermal equilibrium, and largely independent of its material or
geometrical properties, layout of disorder etc~\cite{Fourcade_PRB_1986}. In our
magnetic nanowires, however, disorder is mostly quenched and contributes very
little (see Fig.~3), indicating that the noise in $R$ originates from the
movements of the DWs themselves.

We suggest a mechanism with the help of the schematic shown in the inset of
Fig.~5b, and the Levy-Zhang model of electron scattering within the DWs by
disorder that mixes the spin-up and spin-down channels~\cite{Levy_PRL_1997}. As
the DW moves the scatterers move to the opposite direction with respect to the
DW. Hence the wave function of the electrons within the DW, which depends on
the mistracking of the electron spin to local magnetization, ``see'' a
time-varying layout of the scatterers. This will cause a time-dependent mixing
of the spin-channels, {i.e.} $\rho^\uparrow/\rho^\downarrow$ will fluctuate
with time, leading to fluctuations in the measured $R$. The diffusive kinetics
of the scatterers indicated by the PSDs in Fig.~5c, then implies that the DW
itself moves by diffusive random walk from one the pinning center to the other,
providing the first evidence of such a behavior in magnetic nano-systems. A
distribution function of resistance jumps in these states is difficult to
compare with the theoretical models that associate universal exponents to
distribution of DW displacements~\cite{Durin_PRL_2000,ABBM_JAP_1990}, but we do
observe a power law behavior in such constructions with an exponent of $\approx
2.7$, which is presumably non-universal (inset of Fig.~5c). Nevertheless,
observation of $\alpha \approx 1.5$ in noise can act as a detector of moving DW
inside the nanowire, while static or locally hopping of DWs would lead to
$\alpha$ that is closer to unity.

In conclusion, we have shown that low-frequency fluctuations in electrical
resistance of magnetic nanowires can be a sensitive probe to domain kinetics
under an applied magnetic field. Both noise magnitude and spectral exponent can
detect the number fluctuation and propagation stochasticity of the domain
walls. We find the first evidence of random walk in the propagation of
individual domains along the nanowire at small magnetic fields, that display an
universal kinetic exponent.

We acknowledge the Department of Science and Technology, Government of
India, for funding the work.

\newpage

\begin{figure}[t]
\includegraphics[width=7cm,height=8cm]{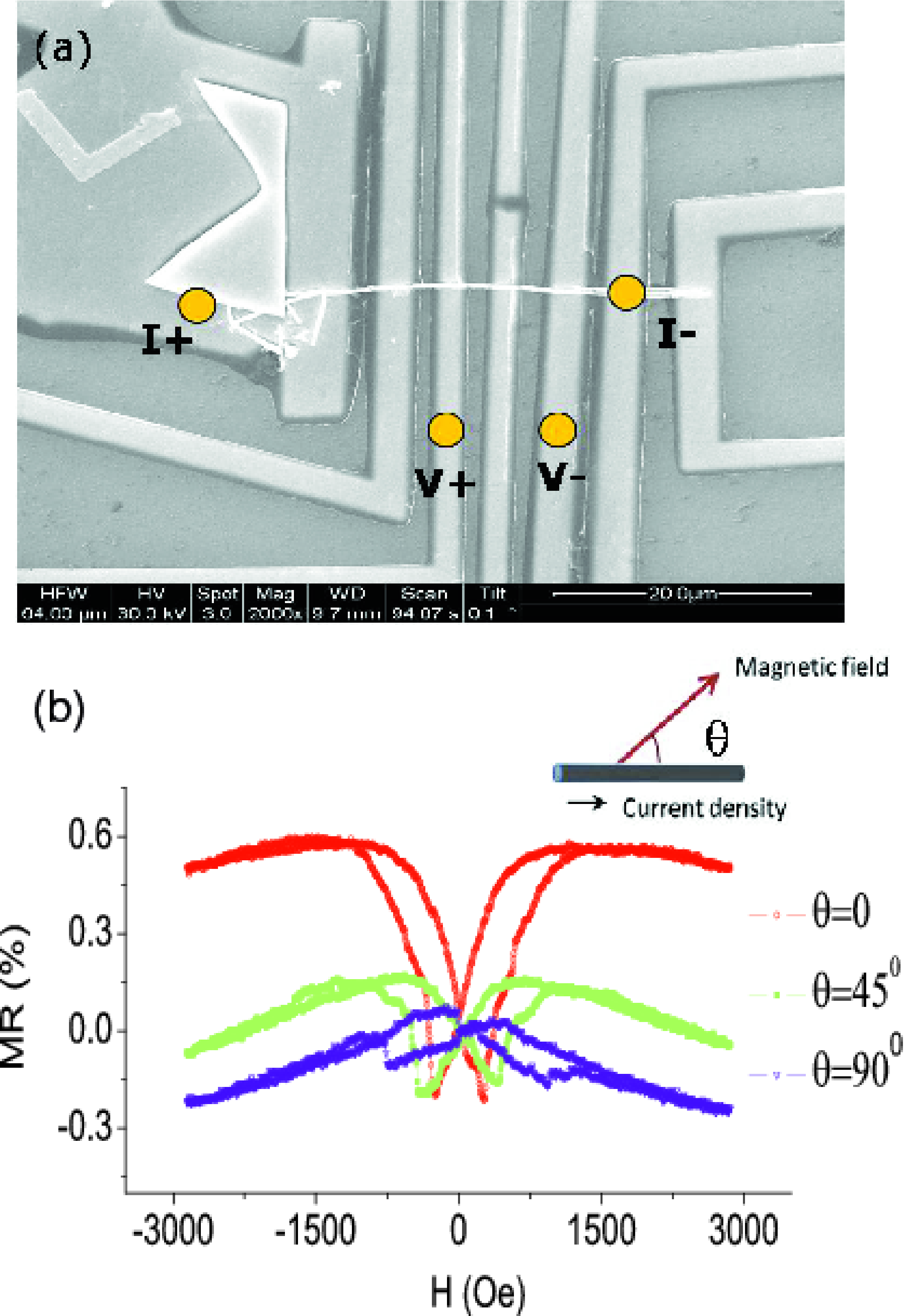}
\vspace{0.8cm}\caption {(color online): (a) Scanning electron
micrograph of the device used in the present experiments. The voltage and
current probes are indicated as $V+/V-$ and $I+/I-$, respectively. (b)
Anisotropic magnetoresistance (AMR) for three different angles between the
current and external magnetic field ($H$)}.
\label{figure1}
\end{figure}

\newpage

\begin{figure}[t]
\includegraphics[width=8cm,height=9cm]{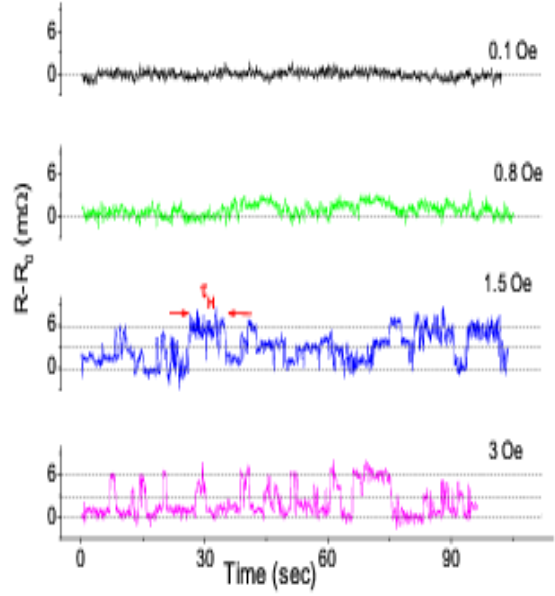}

\vspace{1cm}\caption {(color online): Time variation of resistance
at four different magnetic field applied parallel to the long axis of the
nanowire. The high resistance state for $H = 1.5$~Oe and 3~Oe are indicated by
the arrows and their duration by $\tau_H$. The dashed horizontal lines identify
the discrete resistance states observed in the time traces.}
\label{figure2}
\end{figure}

\newpage

\begin{figure}[t]
\includegraphics[width=11cm,height=10cm]{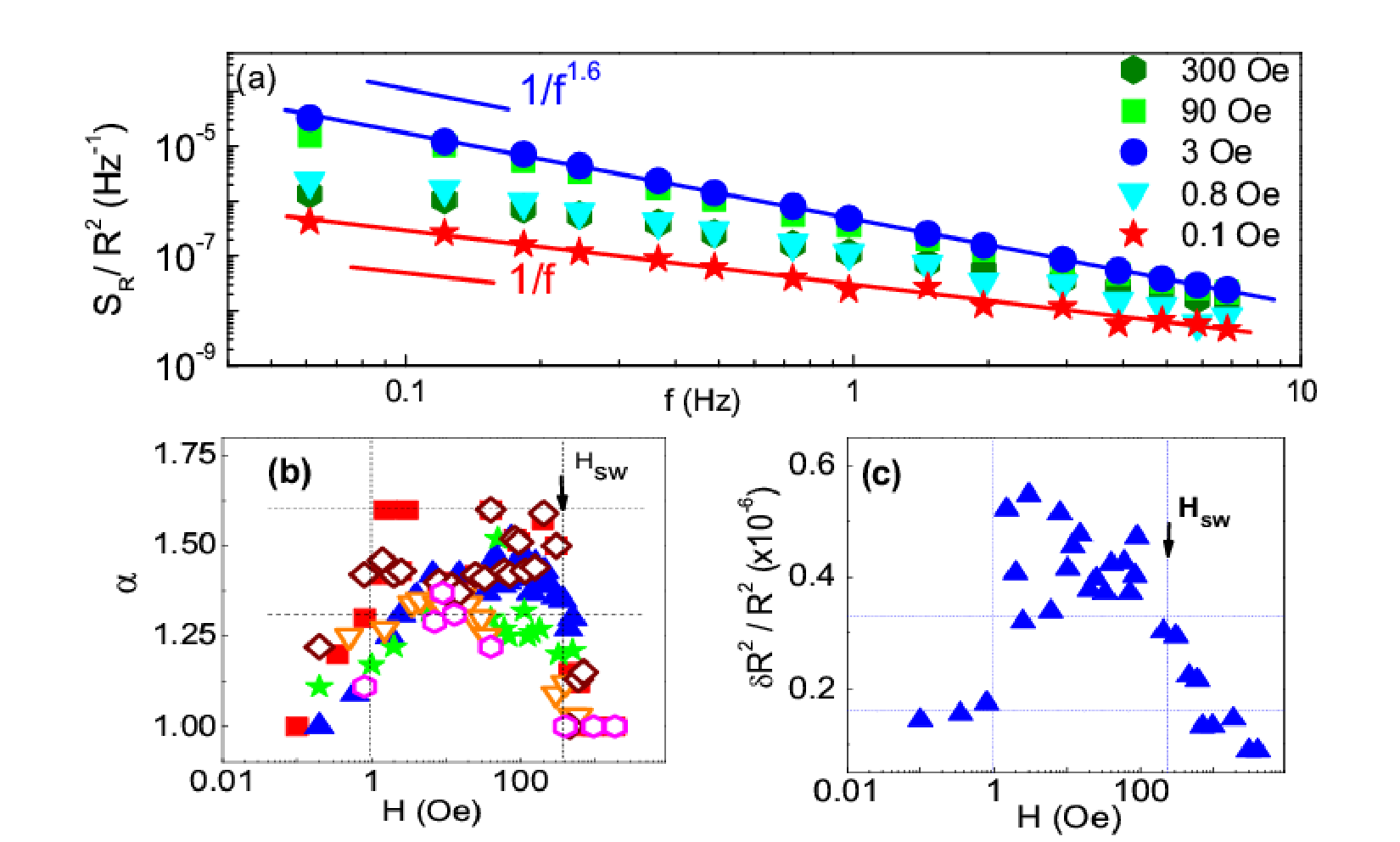}

\vspace{0.2cm}\caption {(color online): (a) Noise power spectral
density (PSD) at different values of $H$. Nonmonotonic variation of (b) the
spectral exponent and (c) the normalized variance in noise. In (b) different
symbols signify different magnetization cycles. The switching field $H_{sw}$
obtained from the AMR measurements is also indicated in (b) and (c).}
\label{figure3}
\end{figure}

\newpage

\begin{figure}[t]
\includegraphics[width=9cm,height=7.5cm]{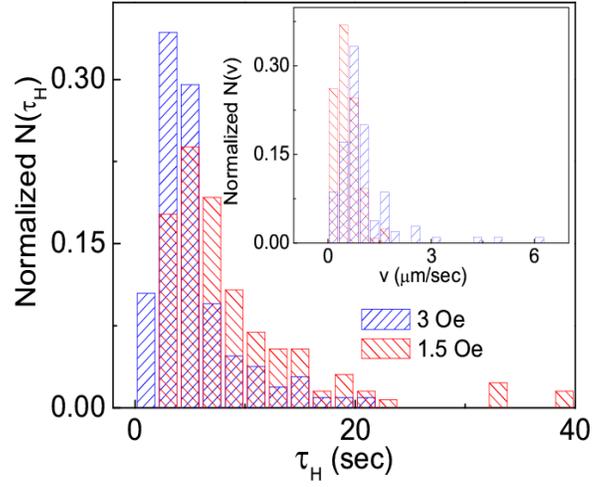}

\vspace{.8cm}\caption {(color online): The distribution of $\tau_H$
for two magnetic fields ($H = 1.5$~Oe and 3~Oe). Inset: Distribution of the
velocity of the domain walls $v = L/\tau_H$, where $L$ is the distance between
the voltage probes, at the same values of $H$.}
\label{figure4}
\end{figure}

\newpage

\begin{figure}[t]
\includegraphics[width=9.5cm,height=10cm]{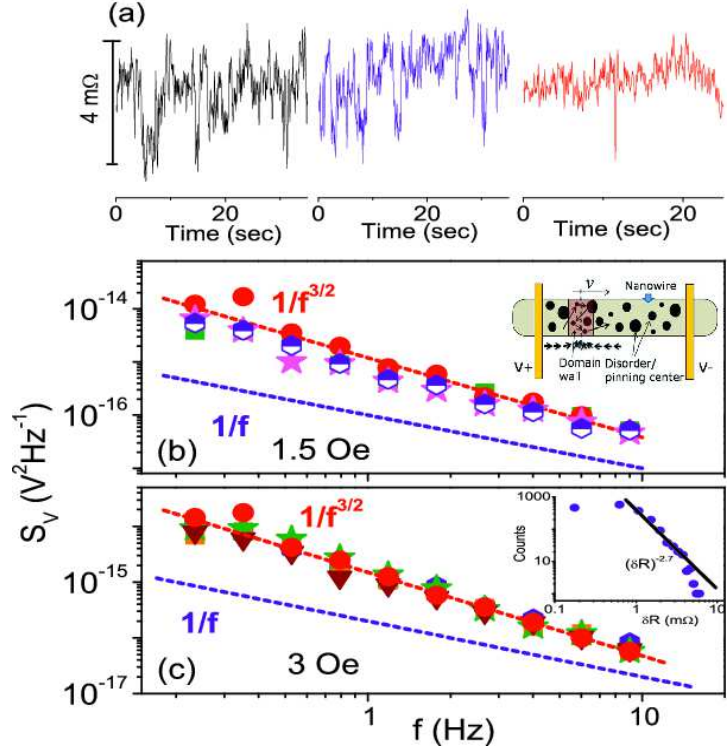}

\vspace{0.2cm}\caption {(color online): (a) Resistance-time behavior
within three high resistance states (see Fig.~2 also). Power spectral density
(PSD) of resistance fluctuations in this states is shown for (b) $H = 1.5$~Oe
and(c) $H = 3$~Oe. Inset of (b): Schematic of electron scattering events within
a domain wall which becomes time dependent as the wall moves between the
voltage probes. Inset of (c): Distribution of resistance jumps (in the
high resistance state) for $H = 3$~Oe.}
\label{figure5}
\end{figure}

\end{document}